\documentclass[a4 paper, 11pt,twoside]{article}
\usepackage{fullpage}                           
\usepackage[lmargin=0.6in,rmargin=0.6in,tmargin=0.6in,headsep=.2in]{geometry}
\usepackage{graphicx}
\usepackage{forloop}
\usepackage{etoolbox}
\usepackage{calc}
\usepackage{wrapfig,lipsum,booktabs} 
\usepackage{xtab} 
\usepackage[dvipsnames]{xcolor}
\usepackage[normalem]{ulem}
\usepackage{sectsty}
\makeatletter
\def\@seccntformat#1{\@ifundefined{#1@cntformat}%
{\csname the#1\endcsname\;}
{\csname #1@cntformat\endcsname}
}
\def\section@cntformat{\thesection.\;} 
\def\subsection@cntformat{\thesubsection.\;} 
\makeatother
\usepackage{hyperref}
\hypersetup{
    colorlinks=true,
    urlcolor=blue,
    linkcolor=black,
    citecolor = black}

\usepackage{fancyhdr}
\fancypagestyle{first}{%
  \fancyhf{}
}
\usepackage{amsmath,amsfonts,amssymb,amsthm,epsfig,epstopdf,url,array}
 \usepackage[retainorgcmds]{IEEEtrantools}
 \usepackage{makeidx,epsfig,lscape}
 \usepackage{xcolor,pict2e}
\usepackage{upgreek}

\newcommand{\atan}{\textnormal{\,atan}}

\newcommand{\dd}{\textnormal{\,d}}

\theoremstyle{definition}

\begin{document}
\thispagestyle{first}
\vspace*{3cm}
{\noindent\huge\bf Hubble Expansion as an Einstein Curvature}\\[1cm]
{\bf\large Marr John H.}\\[0.5cm]
Unit of Computational Science \\
Armstrong Close, Hundon, CO10 8HD, UK \\
Email: john.marr@2from.com \\

{\color{Black}\rule{0.7\textwidth}{2pt}}\\[0.2cm]
{\color{Black}\bf\large Abstract}\\
Extending the spacetime manifold of general relativity (GR) to incorporate the Hubble expansion of space as a specific curvature, generates a modified solution with three additional non-zero Christoffel symbols and a reformulated Ricci tensor and curvature.
The observational consequences of this reformulation are compared with the $\Lambda$CDM model for luminosity distance using the extensive type~Ia supernovae (SNe~1a) data with redshift corrected to the CMB, and for angular diameter distance using the recent baryonic acoustic oscillation (BAO) data.
For the SNe~1a data, the modified GR and $\Lambda$CDM models differ by $^{+0.11}_{-0.15}~\mu_B$~mag. over $z_{cmb}=0.01-1.3$, with overall weighted RMS errors of $\pm0.136$ $\mu_B$~mag for modified GR and $\pm0.151$ $\mu_B$~mag for $\Lambda$CDM respectively.
The BAO measures span a range $z=0.106-2.36$, with weighted RMS errors of $\pm0.034$~Mpc with $H_0=67.6\pm0.25$  for the modified GR model, and $\pm0.085$~Mpc with $H_0=70.0\pm0.25$ for the $\Lambda$CDM model. 
The derived GR metric for this new solution describes both the SNe~1a and the BAO observations with comparable accuracy to the $w'\Lambda$CDM model.
By incorporating the Hubble expansion of space within general relativity as a specific curvature term, these observations may be described without requiring additional parameters for either dark matter or accelerating dark energy.
\vspace{0.5cm}\\
{\color{Black}\bf\large Keywords}\\
Hubble Expansion, General Relativity, Luminosity Distance, Angular Diameter Distance, Dark Mass, Dark Energy
\vspace{0cm}\\
{\color{Black}\rule{0.7\textwidth}{2pt}}

\section{Introduction}
\label{sec:intro}
To the early successes of the precession of the perihelion of Mercury and gravitational bending of star light during a solar eclipse have been added many further observations confirming that General Relativity (GR) well describes the behaviour of masses and photons in a local gravitational field.
Observational data have confirmed without exception that solutions to the general field equations are exact when applied to static or rotating localised gravitational masses, including gravitational redshift \cite{2002Natur.420...51C}, the production of Einstein rings by DM halos \cite{2015ApJ...811..115W}, X-ray emission data in the neighbourhood of black holes \cite{2004A&A...413..861M, 2009ApJ...706..925B, 2014MNRAS.441.3656R}, and the Sunyaev-Zeldovich effect \cite{2012Crowell, refId0}.

In addition to these observations, GR has also been used to analyse an array of observational data using supernovae type~1a (SNe~1a) as "standard candles" and the recent Baryon Acoustic Oscillation (BAO) clustering data as a "standard ruler" for the length scale in cosmology. 
This interpretation using standard GR can only be fully fitted to the data by the addition of a dark mass (DM) component and the incorporation of a hypothetical dark energy, neither of which have been directly observed, and neither of which can be accounted for with current theories of particle physics.

The standard definition of the Hubble expansion coefficient is as velocity per distance (km/s/Mpc), but this may also be interpreted as a velocity per photon travel time, which is dimensionally an acceleration. For $H_0=67.7$~km/s/Mpc, this gives an equivalent value of $H_0\equiv 20.74$~km/s/Myr for photon travel time across the Hubble expansion.
Under GR, any acceleration is equivalent to a curvature, and by considering this expansion to be an additional curvature of space we may extend Einstein's general equation to produce a solution with three additional non-zero Christoffel symbols and a reformulated Ricci tensor and curvature (Section \ref{sec:Geom_curvature}). 
This solution retains the standard components of GR while reducing to the equations of SR as $\Omega_m\rightarrow 0$. 
Sections \ref{sec:dL} and \ref{sec:dA} examine this proposed model by comparing its predictions for luminosity distance (LD) with the extensive apparent magnitude data of supernovae type~1a (SNe~1a), and with a wide range of recently published angular diameter distances from the Baryonic Accoustic Oscillation (BAO) data out to $z=2.36$.  

\section{The FLRW metric}
\label{sec:FLRW}
\begin{figure}[ht]
   \centering
\includegraphics[height=9cm, width=6.5cm]{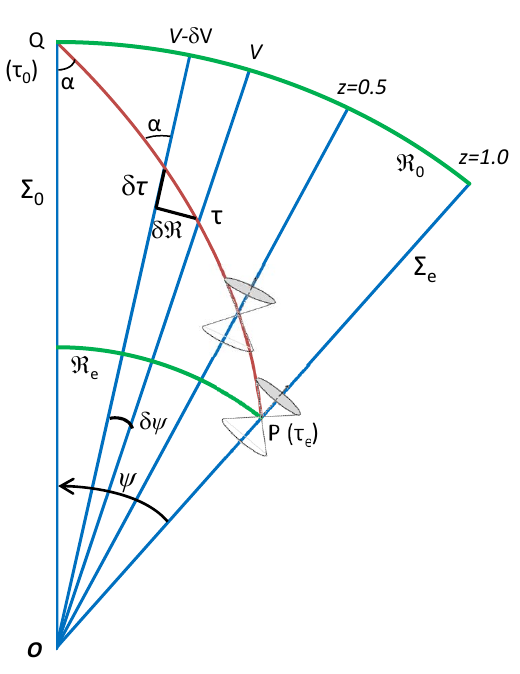} 
   \caption{Geodesic for a photon traversing mass-free space, from frame $\Sigma_e$ moving at velocity $V$ relative to an observer $\Sigma_0$, with a small element of the geodesic $\delta{}S$ for reference frames $V$ and $V-\delta{V}$ rotated through $\delta\psi$, with two local light cones. 
   The  photon path (red line) is a logarithmic spiral, $1+z=\exp\psi$, across diverging galaxies on the complex plane ($c=1\equiv{45}^\circ$).~\cite{2016IJMPC..2750055M}.
   Redshifts referenced to $\Sigma_0$} 
   \label{fig:galaxies}
   \vspace*{8pt}
\end{figure}

Geometrically, the constancy of $c$ for any observer may be represented by the locus of a logarithmic spiral to generate a curve of constant angle to the local time axes \cite{2016IJMPC..2750055M}
(Fig.~\ref{fig:galaxies}). 

This geodesic of SR may be illustrated as a hyperbolic curve crossing diverging imaginary time axes, and is independent of the spatial curvature which is allowed to be flat, spherical or hyperbolic. The Friedmann, Lema\^{i}tre, Robertson, Walker (FLRW) equation allows this expansion curvature of SR to be introduced by the hyperbolic curvature of space as the combined mass-energy of space $\rightarrow0$, which contrasts with observations that show space to be essentially flat.  
In GR, curvature occurs by the distortion of space by gravitational energy, and these gravitational effects on the curvature of the Universe will increase in significance as look-back time extends and temperature and energy densities increase towards the CMB radiation and the early universe at $z\simeq1090$ \cite{refId0}.
The loss of an innate hyperbolic curvature of expansion may be mimicked in GR by introducing extra mass as dark matter (DM) and dark energy as a variable acceleration component, with both components being required and adjusted to match current cosmological observations. 

A model geometry of the evolving Universe may be constructed as a simply connected smooth  Riemannian manifold $R_m$ with metric $g_{\mu\nu}$. 
It is taken as axiomatic that the Universe is homogeneous and isotropic in space, but not in time. 
Of the eight Thurston 3-manifold Riemannian geometries, only three fulfil the criteria of homogeneity and isotropy for the observable Universe: the 3-sphere $S^3$, the 3-D Euclidean space $E^3$, and the 3-D hyperbolic space $H^3$.
Finite volume manifolds with $E^3$ geometry are all compact and have the structure of a Seifert fibre space, remaining invariant under Ricci flow.
$S^3$ manifolds are exactly closed 3-manifolds with a finite fundamental group, and under Ricci flow such manifolds collapse to a point in finite time.
In contrast, manifolds with $H^3$ hyperbolic geometry are open and expand under Ricci flow \cite{2003Milnor}.
Using a Lie group acting on the metric to  compute the Ricci tensor $R_{\mu\nu}$, these manifolds are deformed by Ricci flow as a function of time $t$ and we may then define the geometric evolution equation, $\partial_t d_{ij}=-2R_{ij}$ \cite{2008Perelman}, with normalised Ricci flow given by:

\begin{equation}\label{eq:Ricci_flow}
\partial_t g_{ij}=-2R_{ij}+\frac{2}{3}R g_{ij}\,.
\end{equation}

This is equivalent to a Universe that can be foliated into space-like slices, and spacetime itself may therefore be represented by $\Gamma$--$\mathbb{R}^3$ where $\Gamma$ represents the time direction, with the general form $\dd{s^2}=g_{\mu\nu}\dd{x^\mu}\dd{x^\nu}$ in the standard notation. 
$\mathbb{R}^3$ must be a maximally symmetric space to conform to a homogeneous and isotropic three-manifold, with metric $\dd\sigma^2=\gamma_{ij}\dd{x}^i\dd{x}^j$.
By scaling $t$ such that $g_{00}=-1$ with $\textnormal{c}=1$, we may write the metric as:

\begin{equation} \label{eq:metric2}
\dd{s^2}=-\dd{t^2}+a(t)^2\gamma_{ij}(x)\dd{x^i}\dd{x^j}\,,
\end{equation}

where $\gamma_{ij}$, $x^i$, $x^j$ are the co-moving co-ordinates.

In cosmology, homogeneity and isotropy imply that $\mathbb{R}^3$ has the maximum number of Killing vectors and, with the additional constraint of the metric being torsion-free (the Levi-Civita connection), $\gamma_{ij}$ is the maximally symmetric metric of $\mathbb{R}^3$.
This yields the general solution to Einstein's equation \cite{1970Misner,1993Peebles,2004Carroll} which may be stated in polar coordinates (Eq.~\ref{eq:FLRW2}):

\begin{equation}\label{eq:FLRW2}
\dd{s^2}=-\dd{t^2}+a(t)^2\left[\dd{r^2}+ S_k^2(r)\dd\Omega^2\right]\,,
\end{equation}

\begin{equation}\label{eq:Sk}
\mbox{where~~~}S_k^2(r) \equiv 
\left\{ \begin{array}{lcl}
\Re_0^2\sin^2(r/\Re_0) & \mbox{for} & \Re_0>0  \\
r^2 & \mbox{for} & \Re_0=\infty \\
\Re_0^2\sinh^2(r/\Re_0) & \mbox{for} & \Re_0<0 \nonumber\,,
\end{array}\right.
\end{equation}

\begin{equation}\label{eq:Sk2}
\mbox{or~~~} S_k(r) \equiv \frac{1}{\surd{K}}\sin\left(r\sqrt{K}\right)\nonumber\,,
\end{equation}
and $K=\textnormal{sgn}(\Re_0)/\Re_0^2$ is the curvature.
With $\chi$ as a third angular coordinate, $r=\Re_0\chi$ is the radial distance along the surface of the manifold, $\Re_0$ is the comoving 4-space radius of $\mathbb{R}^3$ at the present epoch, and $\dd\Omega^2=\dd\theta^2+\sin^2\theta\dd\phi^2$ is the angular separation.
The signature {\bf{diag}}$(-,+,+,+)$ defines this as a Pseudo-Riemannian manifold with metric $g_{\mu\nu}$ and spatial metric $g_{ij}$, and $a(t)$ is the scale factor at proper time $t$.
The actual form of $a(t)$ is determined by the curvature of the manifold and the energy tensor of Einstein's field equations, with curvature $K$ (or radius $\Re$), and scale factor $a(t)$ to be determined.
The curvature or shape of the homogeneous hyper-surfaces are defined by the spatial 3-metric $\gamma_{ij}\dd{x}^i\dd{x}^j$ of Eq.~(\ref{eq:metric2}), but the whole dynamics of the Universe are embodied only in the expansion factor, $a(t)$ \cite{1970Misner}.

Just as the surface of a sphere is a curved 2-D manifold embedded in Euclidean 3-space, this manifold is a curved 3-D volume embedded in Euclidean 4-space. 
Measurements on the surface of a 2-D sphere involve a distance and an angle, with the third dimension the implicit radius of the sphere.
For the 3-D volume, $\chi$ is a third angular measure, with the implicit radius $\Re$ now the fourth dimension \cite{doi:10.1063/1.2356388}. 
For an expanding 2-D manifold in 3-D space, time is geometrically a fourth dimension, and---by extension---for the expanding 3-D volume in 4-D space, time must be represented geometrically as a fifth dimension.

To understand physical reality we may invoke geometrical representations, with intrinsic curvature equivalent to embedding in higher dimensions.
This purely geometric dimensionality is distinct from attempts to introduce extra physical dimensions into GR such as by quantum gravity, string theory or loop theory \cite{PhysRevLett.116.071102}, and it must be emphasised that the intrinsic curvature here is a mathematical construct relating the deviation of parallel lines towards or away from each other and does not represent higher physical dimensions. 
With $r$ as the radial coordinate, radial distances are Euclidean but angular distances are not, but if we are only interested in photon redshift distances, $\dd\Omega=0$ and Eq.~(\ref{eq:FLRW2}) is the more useful form of the metric. 
Setting $\dd{s}^2=0$ and $g_{\theta\theta}=g_{\phi\phi}=0$, $\dd{r}$ now represents a radial photon distance from the era of emission $t_e$ to the present epoch at $t_0$, with:

\begin{equation}
R_\gamma = \int\!\dd{r} = \int_{t_0}^t\!\frac{\dd{t}}{a(t)}\,.
\label{eq:R0a}
\end{equation}

$R_\gamma$ is a function of $a(t)$ only, and may be independent of the curvature of the spatial manifold.
Symmetry ensures that proper time for standard clocks at rest relative to the spatial grid is the same rate as the cosmological time $(t)$, making the interval $\dd{t}$ Lorentzian. 
Any coordinate system in which the line element has this form is said to be synchronous because the coordinate time $t$ measures proper time along the lines of constant $x^i$ \cite{1970Misner}.

The substitution $\chi=\sin(r/\Re)$, $\chi=r/\Re$, or $\chi=\sinh(r/\Re)$ into $S_k(r)$ in Eq.~(\ref{eq:FLRW2}) makes $\chi$ a radial coordinate with $\Re$ absorbed into $a(t)$, and now angular distances are Euclidean but radial distances are not (Eq.~\ref{eq:FLRW1}): 

\begin{equation}\label{eq:FLRW1}
\dd{s^2}=-\dd{t^2}+\Re(t)^2\left[\frac{\dd{\chi^2}}{1-k\chi^2}+\chi^2\dd\Omega^2 \right]\,.
\end{equation}

This form is useful for measuring angular distances on a shell of fixed radius ($g_{\chi\chi}=(1-k\chi^2)^{-1}\,,~\dd{\chi}=0$), such as the proper diameters of clusters  or spatial volume for galaxy counts.

\section{The expanding Universe as geometry}
\label{sec:geometry}
Milne described a dust universe expanding with constant relative velocity assigned to each galaxy, and with a mass-energy density sufficiently low that any deceleration could be neglected \cite{1935Milne}. 
Such a universe does not have to be spatially flat, but it does have the property that $\dot{a}(t)= $~constant, and hence $a(t)=a_0 (t/t_0)$, where $a_0$ is the scale factor at the current epoch $t_0$, defined to be $a_0=1$.
Taking Eq.~(\ref{eq:FLRW2}) to be the FLRW metric for the photon path, we may state that $\dd\theta=\dd\phi=0$, and hence $\dd\Omega=0$ and consider only the radial coordinate $\dd{r}$. 
This modified Milne model (M3) is therefore independent of the space curvature: this may be an expanding 3-sphere, a flat 3-sheet, or a 3-saddle. 
What M3 does demand is that the time-like foliation of these 3-spaces is linear; the space itself may be infinite or closed, but will maintain its initial curvature signature whether expanding forever or contracting.

Einstein's first postulate in a system of non-accelerating inertial frames may be summarised as: the velocity of light is constant for any observer, independent of the velocity of the source. 
Interpreting the time coordinate as the imaginary axis has become depreciated, but to do so forces the proper time axis to be a radius of length $\tau=ict$ and allows a graphical interpretation of the interval $S$ to be unvarying under rotation, providing a geometric visualisation to this postulate. 
In Figure~\ref{fig:galaxies}, the infinitesimal geodesic is extended to illustrate the path of photons between galaxies in the uniformly expanding homogeneous, isotropic universe of M3.

This geometrical figure is generated by assuming that: 
(1) observed redshifts represent a true relative motion (whatever the underlying cause); 
(2) galaxies are moving apart with a velocity that is constant over the time period of the observations, generating a set of diverging inertial reference frames in space; 
(3) photons traverse these reference frames at constant velocity $c$ to all local observers, in their local Minkowski space under a Lorentzian transformation;
(4) this is a 'dust' Universe, with no gravitational effects.

Any individual volume of space such as a specific galaxy may be considered stationary within its own reference frame. 
Let us define this reference frame as $\Sigma_0$ for our own local galactic space (Fig.~\ref{fig:galaxies}). 
This neglects small-scale local movements, being a simple representation and first order approximation of an idealised world line for a particle in space, because the components of $v$ are assumed to relate only to local motions that are generally much less than the recessional velocity, and are taken to be zero in most theoretical models of the Universe.

The relative motion of two inertial frames, $\Sigma_0$ and $\Sigma_e$, diverging from a common origin with velocity $v$ may then be viewed as a hyperbolic rotation $\psi$ (the rapidity) of the spacetime coordinates on the imaginary plane (Fig.~\ref{fig:galaxies}). 
This is a Lorentz boost with a rotational 4-matrix {$\Lambda_{\,\nu\,'}^{\mu}$}:
\begin{equation}
x^\mu=\Lambda^{\mu}_{\nu\,'}x^{\nu\,'}
\end{equation}
\begin{center}
$\Lambda^{\mu}_{\nu\,'} =
\left( 
	\begin{array}{cccc} 
    \cosh{\psi} & \sinh{\psi} & 0 & 0  \\
	\sinh{\psi} & \cosh{\psi}  & 0 & 0\\
    0 & 0 & 1 & 0 \\
    0 & 0 & 0 & 1 \\
    \end{array} 
\right)
$
\end{center}

where $\cosh\psi=(1-v^2/c^2)^{-1/2}=\gamma$, $\tanh\psi=v/c=\beta$, and $\sinh\psi=\beta\gamma$, in the standard notation, with {\bf{det}}$\Lambda=+1$. 

Now consider a volume of space receding from us with velocity $v$ as defined by its redshift, with a proper radial distance $\Re_e$ at the time of emission. 
The photon path can now be represented geometrically as a logarithmic spiral on the complex plane ($PQ$ in Fig.~\ref{fig:galaxies}).
It will be noted that $\psi$ is the hyperbolic angle, so the geometry allows $\psi>360^{\circ}$ because $v/c =\tanh{\psi}\rightarrow 1$ as $v\rightarrow c$ and $\psi\rightarrow \infty$, whereas local velocities are represented by real angles with trigonometric functions. 
The scale is chosen by convention such that $\alpha=45^\circ$ with $c=1$, hence the maximum angle in the local frame of reference corresponds to the standard light cone with $\atan(1)=45^\circ$.
Although the spatial component of the M3 model may have curvature, M3 has no matter density and Fig.~\ref{fig:galaxies} is therefore geometrically flat as a consequence of the linear relationship between the radial and time axes. 

For a photon, $\delta{}S=0$ (null geodesic for photon); it then follows that $\delta{}\Re^2=c^2\delta t^2$, or $\delta{}\Re=\pm c\delta{}t$, where the sign represents an incoming or outgoing photon. But $\delta{}\Re=ct\delta{}\psi$, thus $\delta{}t/t=\mp \delta\psi$.
Using $-\delta\psi$ for the incoming photon and integrating:

\begin{equation}
\int_{t_e}^{t_0}\frac{\textnormal{d}{t}}{t}=\int_\psi^0-\textnormal{d}\psi\,.
\label{eq:Int_dt}
\end{equation}

\begin{equation}
\textnormal{i.e. } \ln{(t_0/{t_e})}=\psi \textnormal{ or } t_0/{t_e}=e^{\psi}\,.
\label{eq:log_t}
\end{equation}

Although all diverging world lines are equivalent and will ``see'' photons intercepting and leaving them at velocity $c$, the source lines are Doppler red-shifted with a wavelength of emission $\lambda_e$ in $\Sigma_e$, and a wavelength at observation $\lambda_0$. 
Redshift is defined as:

\begin{equation}
z=\frac{\lambda_0-\lambda_e}{\lambda_e}=\lambda_0/\lambda_e-1
\label{eq:Doppler}
\end{equation}
and setting $\lambda_e=\Delta{t_e}$, $\lambda_0=\Delta{t_0}$, it is easy to show that
\begin{equation}
    1+z = \Delta{t_0}/\Delta{t_e} = t_0/t_e = e^\psi.
\label{eq:z}
\end{equation}
But $e^\psi=\cosh\psi+\sinh\psi$, hence $1+z=\gamma+\gamma\beta=\gamma(1+\beta)$, which is the relativistic Doppler shift in SR, with $z\rightarrow{\infty}$ as $v\rightarrow{c}$. 

\begin{figure}[ht]
   \centering
   \includegraphics[width=10cm]{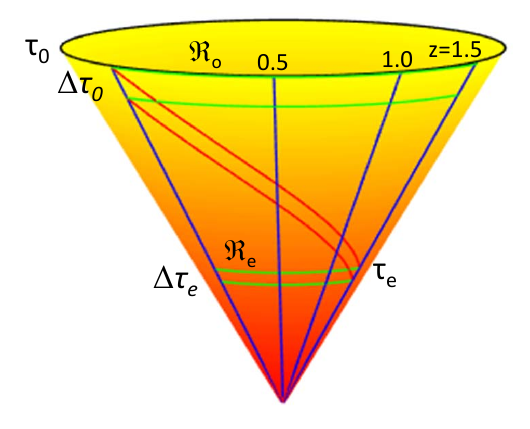}
   \caption{The Milne manifold of Fig.~\ref{fig:galaxies} as a 3--D cone for two photons crossing expanding space, originating at redshift $z = 1.5$ and crossing the paths of galaxies at redshifts $z=1.0$, $z=0.5$, and $z=0$ at constant ($45^\circ$) angles.   The increase in Doppler wavelength ($\Delta\tau_e$ to $\Delta\tau_0$ equivalent to $\lambda_e$ to $\lambda_0$) is visualised in this exaggerated plot.}
\label{fig:MilneCone}
\end{figure}

We may perform a topological transform of the Milne  model of Fig.~\ref{fig:galaxies} into an imaginary 4-cone (Fig.~\ref{fig:MilneCone}) without loss of generality.  
From Eq.~(\ref{eq:z}), $\psi=\log(1+z)$, and the three galaxies represented in Fig.~\ref{fig:MilneCone}, with redshifts of 0.5, 1.0 and 1.5, have corresponding hyperbolic angles of $\psi=23.2^\circ, 39.7^\circ,$ and $52.5^\circ$ respectively.

Despite the appearance of curvature, there is no acceleration ($\dot{a}=$~constant; $\ddot{a}=0$) and this remains a topologically flat figure.
The imaginary proper time axes (e.g.~$\tau_0$ and $\tau_e$) are straight lines that diverge linearly.
Likewise, the radii of curvature round the vertical axis are proportional to $a(t)$, the radial distances on the manifold at constant cosmological (proper) times (e.g. $\Re_0$ and $\Re_e$) are orthogonal functions of $a(t)$ only, and the locus of each photon track is a line of constant angle.

\section{GR as geometry}
\begin{figure}[ht]
   \centering
   \includegraphics[width=10cm]{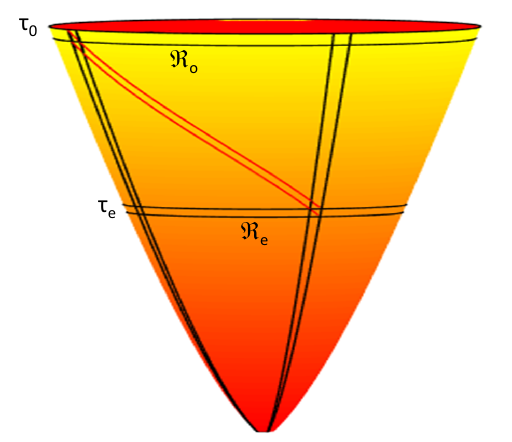}
   \caption{The cone manifold of Fig.~\ref{fig:MilneCone} with curvature of the imaginary time axes by the presence of matter, and two photons crossing the expanding curved space at a constant $45^\circ$ angle.}
\label{fig:GRCone}
\end{figure}

The presence of mass-energy in the Universe introduces a non-linear component to $a(t)$ with consequent curvature of the time axis, and an additional curvature to the path of the photon.
This cannot be displayed on a flat 2-D diagram, but can be demonstrated using the topological transform of Fig.~\ref{fig:MilneCone}.  
The presence of acceleration now introduces curvature to the imaginary $\tau$ coordinate (Fig.~\ref{fig:GRCone}), representing accelerations from gravitational or dark mass and dark energy that may be attractive/positive or negative/repulsive respectively.

\begin{figure} [ht]
   \centering
   \includegraphics[width=10cm]{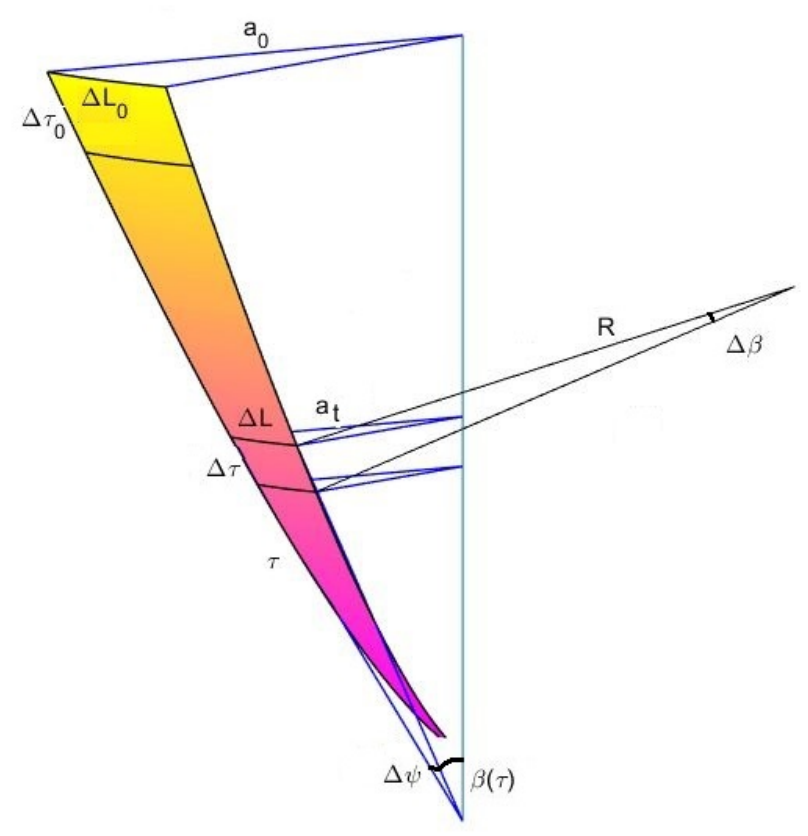}
   \caption{Thin slice of the curved GR manifold (Fig. \ref{fig:GRCone}). 
   $\tau$ is the imaginary time axis. $\Delta\tau_{o}$ is the interval time for the arrival of two photons that were emitted with interval time $\Delta\tau$. 
   $\beta(\tau)=\sinh^{-1}(a/\tau)$ is the Hubble rate of expansion at time $\tau$. 
   $\Delta\beta=\Delta\tau/R$ is rate of change of expansion. 
   $R$ is the mass-energy radius of curvature, considerably foreshortened in this exaggerated plot.} 
\label{fig:petal}
\end{figure}

The manifold of a sphere in 3-space is sufficiently described as a curved two-dimensional surface with time as a third dimension.
Similarly the extra dimensions required to visualise the geometry of an expanding curved spacetime do not represent real dimensions, but are a helpful aid to geometrical visualisation of the manifold. 
Because 3-space with curvature require a 4-dimensional space and the curved time coordinate occupies a further dimension, space-time now exists in 5-space, compacted in Fig.~\ref{fig:GRCone} to a 2-manifold in 3-space. 
Integration of the photon path across this surface may be represented by considering a thin wedge or petal of the time-space manifold in GR (Fig.~\ref{fig:petal}), with the imaginary surface curved by mass-energy as well as by expansion. 

The new radius of curvature is $R(\tau)=1/(\dd\beta/\dd{\tau})$, and this is independent of the spatial curvature, $K$.
In the Milne model, the manifold is flat with $\dd\beta/\dd\tau=0$, and $R=\infty$, and the cone base angle, $\beta_0$, can take any arbitrary value, with $\beta_0=\pi/2$ for Fig.~\ref{fig:galaxies}.
Referring to Fig.~\ref{fig:petal}, the lines of longitude are the imaginary time axes, with $\dd\tau = i\dd{t}$, whilst the lines of latitude represent the spatial component defined by $\dd{L}=\gamma_{ij}(x)\dd{x^i}\dd{x^j}$ (Eq.~\ref{eq:metric2}); $\Delta{L_0}$ is the comoving distance; $\Delta{L}=a(t)\Delta{L_0}$ is the proper distance at time $t$; and the curvature $1/R^2=f(\ddot{a}\,)$ is the acceleration.
It may be noted that---in contrast to a standard $radius~v.~time$ plot with $t$ as the vertical axis---the time axis is here embedded in the manifold.
Unlike Fig.~\ref{fig:MilneCone}, the apex of this cone does not converge onto the vertical axis, but curls round itself as $R\rightarrow0$ and $\ddot{a}\rightarrow\infty$.
The model therefore still requires an inflationary scenario to close the gap and ensure causal connectedness.

\section{Geometry with curvature}
\label{sec:Geom_curvature}
Geometrically, redshift is observed when otherwise parallel curved photon paths diverge from each other, as evidenced in the flat Minkowsky Milne model of Fig.~\ref{fig:MilneCone}.
However, actual spacetime is not flat but has curvature imposed upon it through the presence of gravitational masses, requiring the mathematical interpretation of GR.  
In the modified model of GR presented in this paper, the diverging (redshifted) photons generate a distinct but separate curvature superimposed on the intrinsic curvature of spacetime that can be accounted for as an additional geometrical term in standard GR.

Standard vectors are restricted in the presence of curvature on the spacetime manifold, but we may use Cartan vectors as operators \cite{doi:10.1063/1.2356388}. 
Assign to each particle in the Universe the set of observer-dependent coordinates $x^\mu$.
This represents an invariant line element with proper time $x^0=t$, whose spacetime geometry is encoded in the metric $\dd s^2=g_{\mu\nu}\dd x^{\mu}\dd x^{\nu}$, with space coordinates $x^i=x^i(t)$.
Free particles then move along curved geodesics, with 4-velocity 

\begin{equation}\label{eq:Umu}
U^\mu=\frac{\dd{x}^\mu}{\dd{s}}\,.
\end{equation}

With $t$ as a parameter, the spatial derivatives are the velocity components $U^i=\dd{x^i}/\dd{t}$, and we may introduce the differential operator $\dd/\dd{t}=U^i \partial/\partial{x^i}$, which is the directional derivative along the curve \cite{doi:10.1063/1.2356388}.
The components $U^i$ of the operator now form the local coordinate basis, 

\begin{equation} \label{eq:U}
\overrightarrow{U}=\frac{\dd}{\dd{t}}\,;~~\overrightarrow{e_i}=\frac{\partial}{\partial{x^i}}\,,
\end{equation}

and the basis vectors $\overrightarrow{U}=U^i~\overrightarrow{e_i}$ define the parameterised vector space associated with the point $x^\mu$.

Acceleration may be expressed in terms of Eq.~(\ref{eq:Umu}):

\begin{equation} \label{eq:dU}
\frac{\dd U^{\mu}}{\dd s}=\frac{\partial U^{\mu}}{\partial x^{\alpha}}\frac{\dd x^{\alpha}}{\dd s}=U^{\alpha}\frac{\partial U^{\mu}}{\partial x^{\alpha}}\,.
\end{equation}

The motion is then described by the geodesic equation:

\begin{equation}
\frac{\dd{U^\mu}}{\dd{s}}+\Gamma^\mu_{\alpha\beta}U^\alpha U^\beta=0\,,
\end{equation}

\begin{equation}
\mathrm{i.e.~~} U^{\alpha}\left(\frac{\partial U^{\mu}}{\partial x^{\alpha}}+\Gamma^{\mu}_{\alpha\beta}U^{\beta}\right)\equiv U^{\alpha}\nabla_{\alpha} U^{\mu}=0\,,
\end{equation}

where $\Gamma^{\,\mu}_{\alpha\beta}$ are the Christoffel symbols, defined by:

\begin{equation}
\Gamma^{\,\mu}_{\alpha\beta} = \frac{1}{2}g^{\mu\lambda}(\partial_\alpha g_{\beta\lambda}+\partial_\beta g_{\alpha\lambda}-\partial_\lambda g_{\alpha\beta}) \,.
\end{equation}

\subsection{Curvature of space from the velocity vector}
Parallel transport of a vector is different over different paths.
For redshift observations, we are interested in the parallel transport of photons across an expanding space whose rate of expansion changes with time and distance.
The standard FLRW metric is generally written as a symmetrical function (Eq.~\ref{eq:metric2}), with $\mu,\nu=0\cdots 3$.
However, as demonstrated in Section~\ref{sec:FLRW}, a further curvature term representing the divergence of space may be added to the $R$-axis as a consequence of its expansion.
This requires an additional dimension represented by $z'=\tau\cos{i\psi}$ ($\tau\cosh\psi$ on the imaginary plane), with divergent angle $\psi$ and  $\mu,\nu=0\cdots 4$.

Because $\psi$ is a hyperbolic angle, this geometry allows $\psi>360^{\circ}$, in contrast to local velocities that are represented by real angles and trigonometric functions.
This divergence velocity is not a physical separation velocity in static space, but an observational velocity from the expansion of space itself, and this introduces a new component $\gamma_{\psi\psi} =\tau^2\sinh^2\psi$ to the geodesic equation (Eq.~\ref{eq:metric3}):

\begin{equation}\label{eq:metric3}
\dd{s^2}=-\dd{t^2}+\gamma_{ij}(x)\dd{x^i}\dd{x^j}+\tau^2\sinh^2\psi\dd{\psi}^2\,.
\end{equation}

The time component is $-\dd{t}^2$, the spatial component is $a(t)^2\left[\dd{r^2}+ S_k^2(r)\dd\Omega^2\right]$, and the expansion component is $\tau^2\sinh^2\psi\dd{\psi}^2$.
The corresponding metric to the geodesic, $g_{\mu\nu}$, is:
\begin{equation}
\label{eq:g_mn2}
\left[ 
	\begin{array}{ccccc} 
    -1 & 0 & 0 & 0 & 0 \\
    0 & a(t)^2 & 0 & 0 & 0\\
    0 & 0 & a(t)^2S_k(r)^2 & 0 & 0 \\
    0 & 0 & 0 & a(t)^2S_k(r)^2\sin^2\theta & 0 \\
    0 & 0 & 0 & 0 & \tau^2\sinh^2\psi \\
    \end{array} 
\right]
\end{equation}

\subsection{Christoffel symbols and Ricci curvature} 
This new curvature term introduces an extra component to Eqs.~\ref{eq:Umu} and \ref{eq:dU}, with $\dd{U^\psi}/\dd{s}$ the time rate of change of the curvature of expansion.
The new non-zero Christoffel symbols from Eq.~(\ref{eq:metric3}) are then:

\begin{equation}\label{eq:Christoffel}
\Gamma^{t}_{\psi \psi} = \tau \dot{\tau} \sinh^2\psi;
~ \Gamma^{\psi}_{t \psi} = \Gamma^{\psi}_{\psi t} = \dot{\tau}/\tau;
~ \Gamma^{\psi}_{\psi \psi} = 1/\tanh\psi\,.
\end{equation}

The non-zero components of the Ricci tensor are now:
\begin{eqnarray}
&R_{00}= -3\frac{\ddot{a}}{a} \nonumber \\
&R_{ij}= \left[\frac{\ddot{a}}{a}+2\left(\frac{\dot{a}}{a}\right)^2+2\frac{K}{a^2}+\frac{\dot{a}}{a}\frac{\dot{\tau}}{\tau}\right]g_{ij} \nonumber \\
&R_{\psi\psi}= 3\left(\frac{\dot{a}}{a}\right)\tau\dot{\tau}\sinh^2\psi \nonumber 
\end{eqnarray}
and the Ricci curvature is:

\begin{equation}\label{eq:Rc}
R = 6\left[\frac{\ddot{a}}{a} + \left(\frac{\dot{a}}{a}\right)^2 + \frac{K}{a^2} + \frac{\dot{a}}{a}\frac{\dot{\tau}}{\tau}\right]\,.
\end{equation}

A consequence of these new non-zero Christoffel symbols (Eq.~\ref{eq:Christoffel}) is discussed in Section~\ref{sec:Discussion}.

\subsection{The Einstein equation and mass-density tensor}
The Einstein field equation describes gravity as a manifestation of the curvature of spacetime. 
In particular, the curvature of spacetime is directly related to the energy--stress tensor through the Einstein field equation (Eq.~ \ref{eq:Einstein}): 

\begin{equation}\label{eq:Einstein}
R_{\mu\nu}-\frac{1}{2}Rg_{\mu\nu}=\frac{8\pi G}{c^4}T_{\mu\nu}-\frac{\Lambda}{c^2}g_{\mu\nu}\,,
\end{equation}

where $R_{\mu\nu}$ and $R$ are functions of $g_{\mu\nu}$ and its first two derivatives, and $T_{\mu\nu}$ and $\Lambda$ are the stress-energy tensor and the cosmological expansion parameter respectively \cite{2013RAA....13.1409G}. It may be noted that in the standard solution, the source of curvature is attributed entirely to matter, including dark matter, and $\Lambda$ is a curvature accounted for by dark energy.
For an ideal fluid with mass/unit volume $\rho$ and pressure $P$, the stress-energy tensor in the rest frame of the fluid is $T^{\mu}_{\nu}=(\rho+P)U^\mu U_\nu+P\delta^\mu_\nu$, or:

\begin{equation}
T_{\mu\nu}=(\rho+P)U_\mu U_\nu+Pg_{\mu\nu}\,,
\end{equation}

from which, by assuming symmetry with all off-diagonal components $=0$, setting $c=1$, and using $\dd{a}/\dd\tau=a/\tau$ (Fig.~\ref{fig:petal}) with $\tau^2=-t^2$, we may solve Eq.~(\ref{eq:Einstein}) in terms of $\dot{a}/a$ and $\ddot{a}/a$.

\begin{equation}
\label{eq:FLRW_3}
\left(\frac{\dot{a}}{a}\right)^2 +\frac{K}{a^2}-\frac{1}{t^2}= \frac{8}{3}\pi G\rho+\frac{\Lambda}{3} 
\end{equation}

\begin{equation}
\label{eq:FLRW_4}
\left(\frac{\dot{a}}{a}\right)^2+2\left(\frac{\ddot{a}}{a}\right)+\frac{K}{a^2}-\frac{2}{t^2}=-8\pi G P+\Lambda\,. 
\end{equation}

Using $(\dot{a}/{a})=H(t)$, and eliminating $\dot{a}/a$ from Eqs.~\ref{eq:FLRW_3} and \ref{eq:FLRW_4},

\begin{equation}\label{eq:FLRW_a2a}
H(t)^2=\frac{8}{3}\pi G\rho-\frac{K}{a^2}+\frac{1}{t^2}+\frac{\Lambda}{3}
\end{equation}

\begin{equation}\label{eq:FLRW_a2}
\frac{\ddot{a}}{a}=-\frac{4\pi G}{3}(\rho+3P)+\frac{1}{2t^2}+\frac{\Lambda}{3}\,. 
\vspace*{8pt}
\end{equation}
 
Defining $\rho_c\equiv3H_0^2/8\pi G$ as the critical density of the Universe, and setting Eq.~(\ref{eq:FLRW_a2a}) to the present epoch with $H(t)=H_0$, $a_0=1$, and  $t=T_0$,

\begin{eqnarray}\label{eq:rho_0}
\rho_c=\rho_0-\frac{3K_0}{8\pi G}+\frac{3}{8\pi GT_0^2}+\frac{\Lambda_0}{8\pi G}\,,
\end{eqnarray}
\begin{eqnarray*}\label{eq:omegas}
\textnormal{and defining:~~~}\Omega_m\equiv\frac{8\pi G \rho_0}{3H_0^2}\quad \Omega_K\equiv-\frac{K}{H_0^2}& \\
\Omega_C\equiv\frac{1}{H_0^2T_0^2}\quad \Omega_\Lambda\equiv\frac{\Lambda}{3H_0^2}&\,,
\end{eqnarray*}

Equation~(\ref{eq:rho_0}) may now be rewritten as:
\begin{equation}\label{eq:one}
    1=\Omega_m+\Omega_K+\Omega_c+\Omega_\Lambda\,.
\end{equation}
.
Using $a/a_0=1/(1+z)$, $\dot{a}/a=-\dot{z}/(1+z)$, $\rho=\rho_0 (a_0/a)^{3}$ \cite{1993Peebles}, and the defined density parameters, we may write:

\begin{equation}\label{eq:d_c_int}
d_C=\int_{t_0}^{t_e}\frac{dt}{a(t)}=\int_0^z\left(\frac{a}{\dot{a}}\right)dz=\int_0^z\frac{dz}{H_0 E(z)}
\end{equation}

where $d_C$ is the comoving distance, $\dot{a}/a=H_0E(z)$, and

\begin{equation}
E(z)=[\Omega_m(1+z)^3+\Omega_K(1+z)^2+\Omega_C(1+z)^2+\Omega_\Lambda]^{1/2}\,.
\end{equation}

\subsection{Solutions}
\label{sec:solutions}
Reintroducing $c$, letting $\Omega_\Lambda=0$, and assuming a flat Euclidean Universe with $\Omega_K=0$,  we may state from Eq.~\ref{eq:one}, $\Omega_C=1-\Omega_m$. 
This has an analytical solution in $z$ (Eq.~\ref{eq:soln_z}):
\begin{equation}\label{eq:soln_z}
d_C=\frac{c}{H_0}\frac{1}{\sqrt{1-\Omega_m}} 
    \log\left(\frac{(1+z)\left((1-0.5\Omega_m)+\sqrt{1-\Omega_m}\,\right)}{1+0.5\Omega_m(z-1)+\sqrt{(1-\Omega_m)(1+\Omega_m z)}}\right)
\end{equation}
which reduces to $d_C=(c/H_0)\ln(1+z)$ in the Milne limit $\Omega_m\rightarrow0$.
In Sections~(\ref{sec:dL}) and (\ref{sec:dA}), this derivation for $d_C$ (Eq.~\ref{eq:soln_z}) is compared with luminosity distance measures and the recently extended angular diameter distance measures respectively.

\section{Luminosity Distance}
\label{sec:dL}
\begin{figure*}
   \centering
   \includegraphics[width=\linewidth]{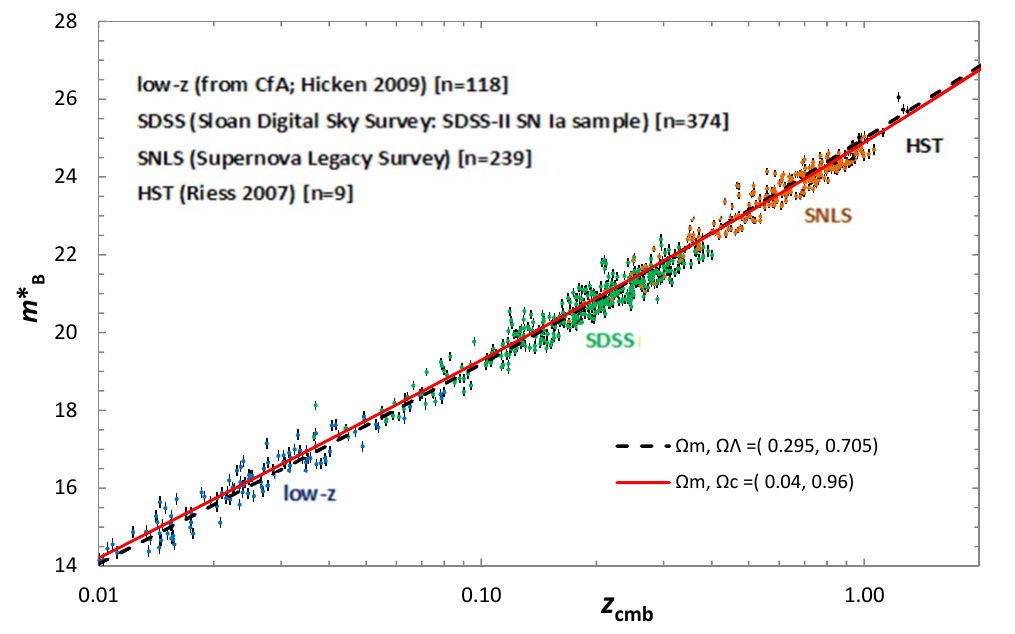}
   \caption{Hubble diagram of the combined sample of 920 SNe\,Ia with the observed peak magnitudes in rest frame $B$-band ($m^*_{\,B}$) \cite{2014A&A...568A..22B} with redshifts corrected to CMB background.
   Overlain are the weighted RMS-minimisation fit for the modified GR model (solid line) and the best-fit $\Lambda$CDM cosmology with $H_0=70$~km~s$^{-1}$~Mpc$^{-1}$, $\Omega_m=0.295$ and $\Omega_\Lambda=0.705$ (dashed line).
   }
\label{fig:Betoule}
\vspace*{8pt}
\end{figure*}

Correlation between the distance modulii derived from the standard $\Lambda$-Cold Dark Matter model ($\Lambda$CDM) and modified gemeral relativity (GR) model was assessed using the extensive type 1a supernovae (SNe 1a) observations \cite{2014A&A...568A..22B}. 
These include SN 1a data for 740 sources \cite{2014A&A...568A..22B}[Table F.3] covering the redshift range $0.01\le z\le 1.3$ and include data from: 
the Supernova Legacy Survey (SNLS) \cite{2006A&A...447...31A}; 
the SDSS SNe survey \cite{2014arXiv1401.3317S}; 
the compilation comprising SNe from SNLS, HST and several nearby experiments 
\cite{2011ApJS..192....1C}; 
photometry of 14 very high redshift ($0.7<z<1.3$) SNe 1a from space-based observations with the HST \cite{2007ApJ...659...98R}; 
and low-z ($z<0.08$) SNe from the photometric data acquired by the Harvard-Smithsonian Center for Astrophysics (CfA3) \cite{2009ApJ...700..331H}.
The corrected apparent brightness parameter $m^*_{\,B}$ for each SN 1a was plotted against its CMB-corrected redshift ($z_{CMB}$) to create the Hubble diagram of Fig.~\ref{fig:Betoule}.
Normalisation depends on the assumed absolute magnitude of the SNe and $H_0$; varying either is equivalent to sliding the curves vertically. 
\begin{figure}
   \centering
   \includegraphics[width=\linewidth]{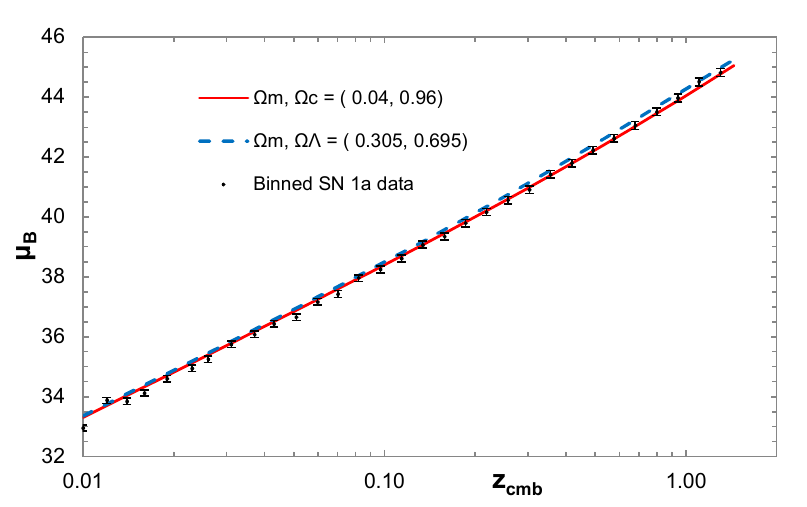}
   \caption{
   Hubble diagram of 920 SNe\,Ia binned logarithmically in $z_{cmb}$, with mean errors and corrected distance modulii $\mu_B$. 
   Overlain are the unweighted least-squares fit for the modified GR model (solid line; RMS error $\pm0.136 \mu_B$~mag) and the best-fit $w$CDM cosmology with $\Omega_m=0.305$, $\Omega_\Lambda=0.695$ (dashed line; RMS error $\pm0.151 \mu_B$~mag). Data from \cite{2014A&A...568A..22B} Table F.1.
   }
\label{fig:Binned}
\end{figure}

Betoule et al \cite{2014A&A...568A..22B} fitted a $\Lambda$CDM cosmology to the SNe measurements by assuming an unperturbed FLRW geometry \cite{1988ARA&A..26..561S}, using a fixed fiducial value of $H_0 = 70$~km s$^{-1}$ Mpc$^{-1}$ ($M_B=-19.12\pm0.05$) to obtain a best fit value for $\Omega_m$ of $0.295\pm0.034$, with $\Omega_\Lambda = 0.705$ (dashed line).
The modified GR model curve (solid line) was fitted by weighted RMS-minimisation to the full data set assuming $\Omega_m=0.04$ as the best current assessment of the mean total observed baryonic density of the Universe, and is comparable to that for the $\Lambda$CDM model (weighted RMS $\pm0.016$ and $\pm0.017$ respectively).
Their $\Lambda$CDM model is 0.15~mag fainter than the modified GR model at $z_{cmb}=1.0$, and the two curves differ by $^{+0.11}_{-0.15}~m^*_{\,B}$~mag over the range $0.01<z<1.3$.

Betoule et al \cite{2014A&A...568A..22B} made a substantial effort to correct the distance modulus for each individual SN, using a parameter ($X_1$) for time stretching of the light-curve, and a colour-correction parameter ($C$) for the supernova colour at maximum brightness \cite{1998A&A...331..815T}.
Using a corrected distance modulus $\mu_B=m^*_{\,B}-(M^*_{\,B}-\alpha X_1+\beta C)$, the resultant plots had less scatter than the raw $m^*_B$ data and became progressively fainter than the $\Lambda$CDM curve with increasing $z_{cmb}$ (Fig.~\ref{fig:Binned}).
To correct for this, they considered three alternatives to the basic $\Lambda$CDM model: 
(a) a non-zero spatial curvature, $\Omega_k$; 
(b) a $w$CDM model with an arbitrary constant equation of state for the dark energy 
with the parameter $w$ equivalent to the jerk parameter of Riess et al \cite{2004ApJ...607..665R}; 
(c) a time-dependent equation of state with a third-order term equivalent to the snap parameter, $w'$ \cite{2004ApJ...607..665R}.  
They concluded that the best overall fit was to a flat universe with typical $\Omega_k\simeq{0.002}\pm0.003$, and a $w$CDM model, with $w=-1.018\pm0.057$ (stat+sys), and with these corrections their $w$CDM curve overlays the binned plots at the faint end (Fig.~\ref{fig:Binned}).
The modified GR model was normalised to the standard model at $z = 0.01$. 
The overall unweighted RMS errors remain comparable for the $w$CDM and modified GR models, being $\pm0.151$ and $\pm0.136$~$\mu_B$~mag. respectively, differing by $^{+0.00}_{-0.24}~\mu_B$~mag. over the range $z_{cmb}=0.01-1.3$.

\subsection{Intrinsic errors to the SNe data set}
Evidence for the presence of dark energy comes most strongly from the measurement of galaxy distances using SNe~1a markers. 
This result is based on the assumption that the corrected brightness of supernovae do not evolve with look-back time, but this assumption has been challenged by a number of more recent observations. 
Shanks et al have suggested that metallicity-dependence of the Cepheid P-L relation is stronger than expected, decreasing the value of $H_0$ \cite{2002ASPC..283..274S}. 
This may impact on the corrected Cepheid distances to galaxies with SNe, suggesting that the supernova peak luminosity is fainter in metal poor galaxies, and the evidence for a non-zero cosmological constant from the SNe 1a Hubble Diagram may be subject to corrections for metallicity which are as big as the effects of cosmology.
Meyers et al studied the properties of 17 SNe~1a at high redshift ($z>0.9$) in early-type galaxies, confirming that the SNe in these hosts brighten and fade more quickly than those hosted by late-type galaxies and may be related to the mass of the host galaxy, although the errors from this were likely to be $<0.06$ mag \cite{2012ApJ...750....1M}.
Other recent studies have shown that the standardised brightness of SN~Ia correlates with host morphology, host mass, and local star formation rate (SFR) \cite{Kang_2020}.
These studies suggest that much of the $H_0$ residual used to support dark energy appears to be affected by SN luminosity evolution. 
Thus, whilst modern observations remain impressive in their extent, detail and number, there remains an overall element of error of $\simeq\pm0.15$~mag, which is within the error range of fitting the $\Lambda$CDM curves and the GR model to the SN data. 

\section{Angular Diameter Distance}
\label{sec:dA}
Angular diameter distance $d_A$ is defined for an object of known proper size $D$, that subtends an angle $\phi$ to the observer such that 

\begin{equation}\label{eq:16}
d_A=D/\phi \,.
\end{equation}

Experimental verification for $d_A$ is notoriously difficult because of the unknown evolution of galaxies, clusters and quasars \cite{2006ApJ...647...25B,doi:10.1117/12.552473,2015ApJS..216...27B}, but recent work using the phenomenon of baryonic acoustic oscillation (BAO) as a suitable measuring rod has enabled measurements of $d_A$ with considerable accuracy.
The BAO signal is one of the key modern methods for measuring the expansion history. 
The BAO arose because the coupling of baryons and photons in the early Universe allowed acoustic oscillations to develop that led to anisotropies of the cosmic microwave background (CMB) radiation and a rich structure in the distribution of matter \cite{2005ApJ...631....1G,2012MNRAS.427.3435A}.
The acoustic scale length ($r_S$) can be computed as the comoving distance that the sound waves could travel from the Big Bang until recombination. The imprint left by these sound waves provides a feature of known size in the late-time clustering of matter and galaxies, and by measuring this acoustic scale at a variety of redshifts, one can infer $d_A(z)$ and $H(z)$. 

BAO is independent of galactic evolution, with the points of $D$ fixed on the surface of the space-like sphere defined by the proper radius $\Re_e$ (Figs.~\ref{fig:MilneCone}, \ref{fig:GRCone}), where we identify $\Re_e$ with the angular size distance. 
This may be used with the standard expression for $d_A$ \cite{2000astro.ph..1419H,2006ApJ...647...25B} in terms of $d_C$ from equation (\ref{eq:d_c_int}):
\begin{equation}\label{eq:dA}
d_A=\frac{d_C}{(1+z)}\,.
\end{equation}

Determination of $r_S$ comes from the the matter-to-radiation ratio and the baryon-to-photon ratio, both of which are well measured by the relative heights of the acoustic peaks in the CMB anisotropy power spectrum \cite{1998ApJ...504L..57E,2013PhR...530...87W}.
Both cosmological perturbation theory and numerical simulations suggest that this feature is stable to better than 1\% accuracy, making it an excellent standard ruler. 
The propagation distance of the acoustic waves becomes a characteristic comoving scale fixed by the recombination time of the Universe after approximately 379,000 years, at a redshift of $z\simeq 1089$ \cite{1970ApJ...162..815P,1970Ap&SS...7....3S,1978SvA....22..523D}.
Eisenstein et al provide a discussion of the acoustic signal in configuration space \cite{2007ApJ...664..660E}, and reviews of BAO as a probe of dark energy \cite{2008PhT....61d..44E}. 
The acoustic scale is expressed in absolute units (Mpc) rather than $h^{-1}$~Mpc, and is imprinted on very large scales ($\sim150$~Mpc) thereby being relatively insensitive to small scale astrophysical processes, making BAO experiments less sensitive to this type of systematic error \cite{2013PhR...530...87W}. 
\begin{figure}[ht]
   \centering
   \includegraphics[width=\linewidth]{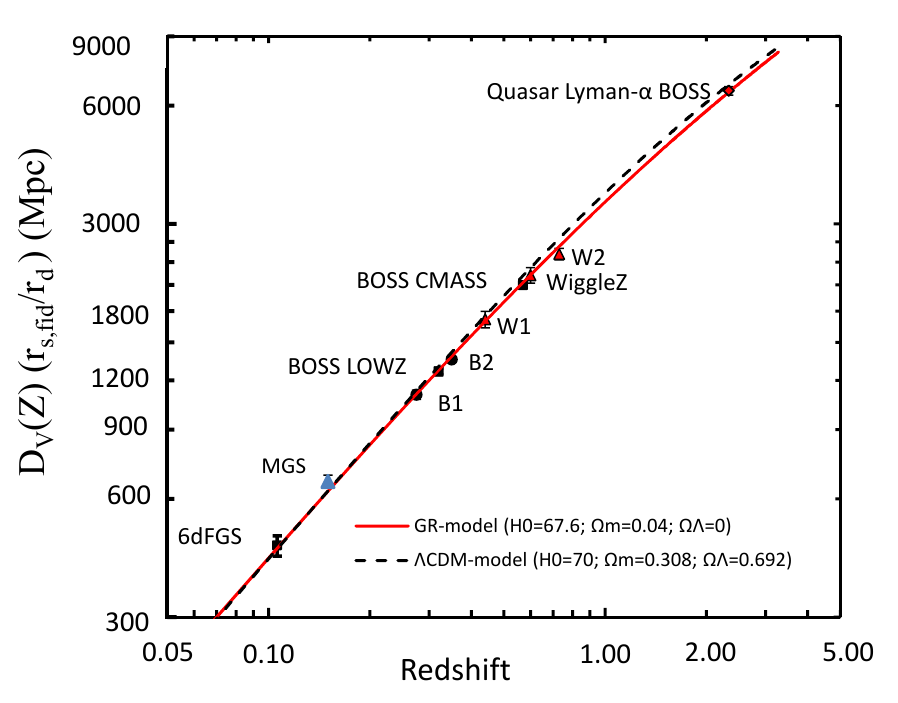}
   \caption{    
   A plot of the distance-redshift relation from the spectroscopic data BAO measurements and quasar Lyman-$\alpha$ BOSS, plotting $D_V(z)(r_{s,fid}/r_d)$ (Table~\ref{table:BAO}).  
   Overlain are the modified GR model fitted by weighted RMS-minimisation to $H_0=67.6\pm0.25$ with $\Omega_m=0.04, \Omega_C=0.96$ (red solid line) and the best-fitting flat $\Lambda$CDM 1-$\sigma$ prediction from WMAP under the assumption of a flat universe with a cosmological constant ($\Omega_m=0.308$; $\Omega_\Lambda=0.692$) \cite{2011ApJS..192...18K,2012MNRAS.427.3435A} (dashed line).
   }
   \label{fig:BAO}
\end{figure}

\begin{table} [ht]
\caption{Parameters from the BAO surveys}
\begin{center}
\begin{tabular}{l c c c}
\hline
\noalign{\smallskip}
Survey&$z$&$D_V (r_d/r_{d,fid})$&Ref \\
 & & (Mpc)\\
\hline
\noalign{\smallskip}
6dFGS&0.106 & 456$\pm27$& \cite{2011MNRAS.416.3017B} \\
MGS&0.15&664$\pm25$& \cite{2015MNRAS.449..835R} \\
BOSS (B1)&0.275&1104$\pm30$& \cite{2010MNRAS.401.2148P} \\
BOSS LowZ&0.32  &1264$\pm$25& \cite{2014MNRAS.441...24A,2016MNRAS.457.1770C} \\
BOSS (B2)&0.35&1356$\pm25$& \cite{2012MNRAS.427.2132P,2013MNRAS.431.2834X} \\
WiggleZ (W1)&0.44&1716$\pm83$& \cite{2014MNRAS.441.3524K} \\
CMASS&0.57&2056$\pm$20& \cite{2014MNRAS.441...24A,2016MNRAS.457.1770C} \\
WiggleZ (W2)&0.6&2221$\pm101$& \cite{2014MNRAS.441.3524K} \\
WiggleZ&0.73&2516$\pm86$& \cite{2014MNRAS.441.3524K} \\
Lyman-$\alpha$ forest&2.36& 6474$\pm$163& \cite{2014JCAP...05..027F} \\
\hline
\end{tabular}
\end{center}
\label{table:BAO}
\end{table}

Figure \ref{fig:BAO} combines the BAO results from a number of sources using spectroscopic data sets, and the quasar Lyman-$\alpha$ results from the SDSS-III Baryon Oscillation Spectroscopic Survey (BOSS).
The volume $D_V(z)$ corresponds to the peak position for an isotropic distribution of galaxy pairs and the 2-point isotropic clustering strength $\xi(z)$ of the observations, computed using 
$D_V\equiv[d_A^{\,2}cz/H(z) (1+z)^2]^{1/3} $ to convert the line-of-sight distance into an equivalent transverse length scale, where $d_A$ is the angular diameter distance and $H(z)$ is the Hubble parameter in the appropriate model.
As the BAO method actually measures $D_V/r_d$, this quantity was multiplied by the fiducial scale length $r_{s,fid}$ to restore a distance \cite{2012MNRAS.427.3435A,2005ApJ...633..560E}. 


Included are the acoustic peak detection from the 6dF Galaxy Survey at $z = 0.106$ \cite{2011MNRAS.416.3017B}; 
the MGS survey at $z=0.15$ \cite{2015MNRAS.449..835R};
a combination of Sloan Digital Sky Survey (SDSS)-II DR7 LRG and main sample galaxies combined with the 2dF data (B1) at $z=0.275$ \cite{2010MNRAS.401.2148P}; 
the BOSS CMASS measurements at $z=0.32$ and $z=0.57$ \cite{2014MNRAS.441...24A,2016MNRAS.457.1770C}; 
the SDSS-II LRG (B2) measurement at $z = 0.35$ using reconstruction to sharpen the precision of the BAO measurement \cite{2012MNRAS.427.2132P,2013MNRAS.431.2834X}; 
and the WiggleZ measurement of three partially covariant data sets at $z = 0.44$, $0.6$, and $0.73$ \cite{2014MNRAS.441.3524K}. 
The published values for $D_V(z)$ are presented in Table~\ref{table:BAO}.
Font-Ribera et al \cite{2014JCAP...05..027F} measured the large-scale cross-correlation of quasars with the Lyman-$\alpha$ forest absorption, using over 164,000 quasars from DR11 of the SDSS-III BOSS. 
Their result was an absolute measure of $d_A=1590\pm60$~Mpc at $z=2.36$, equivalent to $D_V=6474\pm163$~($r_d/r_{s,fid}$)~Mpc, with $r_d = 147.49$~Mpc. 

The data of Fig.~\ref{fig:BAO} are overlain with the best-fit curves for the two models. 
The solid curve is the modified GR model with $\Omega_m=0.04$, $\Omega_C=0.96$, and the dashed line is the $\Lambda$CDM prediction from WMAP under the assumption of a flat universe with a cosmological constant using Planck Collaboration data ($\Omega_m=0.308\pm0.012$; $\Omega_\Lambda=0.692\pm0.012$; $\Omega_K=0$) \cite{refId0}.

As in Section~\ref{sec:dL}, changing $H_0$ slides the curves up or down the vertical axis, but does not alter the shapes of the curves which were fitted by weighted RMS~minimisation against the combined BAO samples of Table~\ref{table:BAO}.
For the modified GR model, $H_0=67.6\pm0.25$ with weighted RMS errors of $\pm0.034$~Mpc in good concordance with the most recent Planck results of $H_0=67.4\pm0.5$ \cite{refId0}, rather than the high value of Riess ($H_0=73.24\pm1.7$) \cite{2018ApJ...861..126R, 2019NatAs...3..891V}.
For the $\Lambda$CDM model, $H_0=70.0\pm0.25$ with  weighted RMS errors $\pm0.085$~Mpc which is intermediate between the two extremes. 
The uncertainties in the two lines come largely from uncertainties in $\Omega_m h^2$ but, as with the luminosity distance measures, the standard model can be improved with non-linear parameters added to $\Omega_\Lambda$ in a $w'$-CDM model. 

\section{Discussion} 
\label{sec:Discussion}
While the nature of dark matter and dark energy remain elusive, several alternative theories to standard GR have emerged \cite{2019BAAS...51c..97S, 2022Astro...1....1A}.
Recently published work following the observation of gravitational waves from the binary neutron star GW170817 \cite{2017arXiv171106843T} have, however, determined $c_g=c\pm10^{-15}$ with sufficient accuracy to eliminate several gravitational theories that predict an anomalous $c_g$ propagation speed \cite{2017PhRvL.119y1303S,2017PhLB..765..382L}, such as some MOND-like gravities including Tensor-Vector-Scalar gravity (TeVeS), Hern\'{a}ndez forms, Einstein-Aether, Generalised Proca and Ho\u{r}ava gravity \cite{2017PhRvL.119y1304E}, and scale invariance as an alternative to dark energy \cite{2017ApJ...834..194M}.

We explore the consequences of considering Hubble expansion as a distinct curvature of space, across which photons move from distant galaxies until observed locally on Earth.
The standard definition of the Hubble expansion coefficient as velocity per distance (km/s/Mpc) can be reformulated as a velocity per photon travel time.
Dimensionally, this is an acceleration with $H_0\equiv 20.74$~km/s/Myr as the photon transit time across an expanding Universe (for $H_0=67.7$~km/s/Mpc).
The first postulate of special relativity (SR) states that the velocity of light $c$ is constant for all observers in their local reference frame, leading to a central tenet of GR: that it is always valid to choose a coordinate system that is locally Minkowskian.
For this to remain true across an expanding Universe with time as one of the axes, the photon path must curve geometrically in a logarithmic spiral \cite{2016IJMPC..2750055M}, and this curvature has the dimensions of an acceleration across the expanding space, represented by an additional curvature term in Einstein's general equation for GR.

The extension to GR presented in this paper incorporates both kinematic and gravitational components,
with parallel transport along the photon path and rotation across curved diverging time lines. 
Non-zero Christoffel symbols arise with any acceleration, whether from a gravitational field, or by the action of fields other than those associated with gravitational mass, or by curvilinear motion \cite{1972Weinberg}.
The emergence of new non-zero Christoffel symbols (Eq.~\ref{eq:Christoffel}) supports the presence of curvilinear motion imposed on the red-shifted photons by the expansion of space ($\Omega_C$) that is distinct from the curvature of space by the presence of mass ($\Omega_M$) or any intrinsic curvature within space itself ($\Omega_K$).

By considering GR as a geometrical manifold with an imaginary time-axis, time-separated photon paths trace out a thin ribbon that everywhere subtends an angle of $45^\circ$ to the expanding time axes, this being the locally Minkowskian metric.
In a static universe with no relative velocity between emitter and receiver, this is a plane ribbon-like Euclidean quadrilateral with parallel time-lines and parallel photon paths, and it retains this form when wrapped round a cylinder.
In the Milne SR model the relative velocity of emitter and receiver cause an intrinsic curvature of space on which this ribbon is curved and stretched to produce the observed redshift whilst retaining the feature of constant $c$ to every observer in the path of the photon stream (Fig.~\ref{fig:galaxies}) \cite{2016IJMPC..2750055M}. 
This curvature can, however, be wrapped without distortion round a uniform cone (Fig. \ref{fig:MilneCone}).

In contrast, the presence of mass-energy ($\rho_0$ and $P$) generates an additional curvature and twist in the ribbon (Figs.~\ref{fig:GRCone}, \ref{fig:petal}) requiring Einstein's equations of GR, generally solved using the standard FLRW metric.
Assuming spatial curvature to be zero, the observed matter in the Universe is insufficient in this model to account for the measured SN redshifts and requires the inclusion of an additional dark-matter component, while to conform to the more detailed SNe~1a measurements an additional dark-energy acceleration, $\Lambda$, is included, mathematically equivalent to a gravitationally repulsive negative mass \cite{1999ApJ...517..565P}. 
Deeper and more detailed SNe~1a measurements have required second and third order refinements to $\Lambda$, with jerk ($w$) and snap ($w'$) parameters \cite{2014A&A...568A..22B, 2004ApJ...607..665R}.

\section{Conclusions} 
\label{sec:Conclusions}
The introduction of an independent curvature term from the expansion of space generates a magnitude-redshift curve that well matches current SNe~1a observations out to $z=1.3$, with $\rho_m$ representing observable baryonic mass.
For the SNe~1a data, the modified GR and $\Lambda$CDM models differ by $^{+0.11}_{-0.15}~\mu_B$~mag. over $z_{cmb}=0.01-1.3$, with overall weighted RMS errors of $\pm0.136$ $\mu_B$~mag for modified GR and $\pm0.151$ $\mu_B$~mag for $\Lambda$CDM respectively.
BAO measurements for angular diameter distances also give an excellent fit from low-$z$ out to $z=2.36$ without requiring additional or arbitrary parameters.
The combined BAO samples of Fig.~\ref{fig:BAO} (Table~\ref{table:BAO}) have weighted RMS errors of $\pm0.034$~Mpc for the modified GR model, and $\pm0.085$~Mpc for the $\Lambda$CDM model with $H_0=67.6\pm0.25$, in good concordance with the recent Planck results \cite{refId0}.
On both the SNe~1a and the BAO data, the modified GR model is comparable to the bast current $w$'CDM models \cite{2021JCAP...01..006D}.

The nature of and theoretical basis for dark energy currently remains as elusive as quintescence or the luminiferous ether \cite{1998PhRvL..80.1582C}.
Although DM may still be required within galaxies to account for galactic rotation, gravitational lensing, and the motion of clusters, there is still neither theoretical nor direct observational support for it \cite{2021NatPh..17.1396A}.
By treating the Hubble expansion of space as an independent curvature in general relativity, the equations of GR can accommodate a scenario in which the observations of SNe~1a and BAO may be described without requiring  additional parameters for DM or accelerating dark energy.



\section{Declarations}
The author declares no competing interests.

\section*{Acknowledgement}
My sincere thanks to the anonymous reviewers of this article for their constructive and helpful comments.

\bibliographystyle{sn-mathphys}
\bibliography{references5}
\end{document}